\documentclass[11pt,a4paper]{article}
\usepackage[margin=27mm]{geometry}
\usepackage{amsmath,amssymb,amsthm}
\usepackage[T1]{fontenc}
\usepackage{lmodern}
\usepackage[hidelinks]{hyperref}
\usepackage{simplewick}
\newtheorem{thm}{Theorem}
\newtheorem{lem}{Lemma}
\newenvironment{pf}{\par\noindent\textit{Proof.}\ }{\hfill$\square$\par}
\title{A Fa\`a di Bruno Convolution Identity in Parafermion Operator Products}
\author{Samuel Oliver Kault\\[4pt]
\small School of Mathematics and Physics\\
\small The University of Queensland\\
\small Brisbane, Queensland, Australia}
\date{}
\begin{document}
\maketitle
\begin{abstract}
We give a detailed combinatorial proof of a Fa\`a di Bruno
convolution identity used in parafermion operator-product
calculations. The identity establishes the cancellation of
potentially more singular terms and shows that the leading
contribution combines the polynomial indices additively.
The proof, originating in the author's doctoral thesis,
provides a general derivation of an identity stated in
earlier parafermion literature.
\end{abstract}

\section{Introduction}

Fa\`a di Bruno polynomials arise naturally when differentiating
exponentials of fields.  In particular, products of operators built from
these polynomials lead to finite sums involving products of polynomials
and their derivatives.  In the parafermion setting, a specific
convolution of this kind has an unexpectedly simple leading term:
two polynomial indices combine into their sum.

Parafermionic conformal field theories were developed
in the foundational work of Zamolodchikov and Fateev
\cite{ZF1985}, and subsequently studied by Gepner and
Qiu \cite{Gepner1987}. Early free-field constructions
of parafermionic operators include those of Bilal
\cite{Bilal1989}, Gerasimov et al.~\cite{Gerasimov1990},
and Camino, Ramallo and S\'anchez de Santos
\cite{Camino1998}.

Camino, Ramallo and S\'anchez de Santos state the
Fa\`a di Bruno convolution identity considered here
in Appendix A, equation (A.14), without presenting
a general derivation. A proof was subsequently
developed in the author's 2012 doctoral thesis
\cite{Kault2012}.

The purpose of the present paper is to make that
argument available in the journal literature.
We provide a detailed derivation of the convolution
identity, including the necessary operator-product
calculations and combinatorial identities.
To our knowledge, this is the first journal
publication of a general proof of the identity
stated in \cite[Appendix~A, Eq.~(A.14)]{Camino1998}.

Theorem~\ref{mainresult} states the operator-product result.
After treating the boundary cases directly, we reduce the remaining
case to the convolution identity in
Theorem~\ref{convolutiontheorem}. This identity explains why the
apparently more singular contributions cancel and why the leading
term contains the single polynomial $P_{l+m}$.
Mixed-sign products and the full parafermion algebra lie outside
the scope of this note.

\section{Preliminaries and derivation of the general OPE}

The derivation below follows the approach developed in the
author's doctoral thesis~\cite{Kault2012}. We introduce the
free-boson conventions, point splitting and Fa\`a di Bruno
polynomials needed to derive the general operator-product
expansion (OPE) of two $\mathcal J$ operators. The coefficient
$Y$ obtained in this calculation will then be specialised
to the hypotheses of Theorem~\ref{mainresult}.

\subsection{Free-boson contractions and vertex operators}
Let $\varphi_1,\ldots,\varphi_d$ be mutually orthogonal chiral
free bosons with contraction
\begin{align}
\contraction{}{\varphi_i(z)}{}{\varphi_j(w)}
\varphi_i(z)\varphi_j(w)
=-\delta_{ij}\log(z-w).
\end{align}
Here the arch denotes a contraction, rather than the full
operator product.

We shall use vector notation for linear combinations of
these bosons. Let $B=(B_1,\ldots,B_d)$ be a constant vector
of complex numbers, and define
\begin{align}
B(z)=B_j\varphi_j(z),
\label{wjksejdx}
\end{align}
where we adopt the Einstein summation convention.
A letter without an argument denotes a vector of complex
coefficients, while the same letter with an argument denotes
the corresponding linear combination of bosonic fields.
This convention applies to both Latin and Greek letters.
For example, $\alpha(z)\equiv\sum_j\alpha_j\varphi_j(z)$.

We use the bilinear dot product
$B\cdot C=\sum_j B_jC_j$, without complex conjugation.
By linearity of contractions,
\begin{align}
\contraction{}{B(z)}{}{C(w)}
B(z)C(w)
&=
B_jC_l
\contraction{}{\varphi_j(z)}{}{\varphi_l(w)}
\varphi_j(z)\varphi_l(w)
\nonumber\\
&=-B_jC_l\delta_{jl}\log(z-w)
\nonumber\\
&=-B\cdot C\log(z-w).
\label{vectorcontraction}
\end{align}
Powers of $z-w$ are understood with a consistent local
choice of logarithm.

The exponential of a bosonic field requires some care, since
products of fields evaluated at the same point may be singular.
We therefore introduce the \emph{vertex operator} associated
with a bosonic field $\phi(z)$, defined by the normal-ordered
exponential \cite{DiFrancesco1997},
\begin{align}
V_a(z)
&=:e^{a\phi(z)}:
=\sum_{n=0}^{\infty}\frac{a^n}{n!}:\phi(z)^n:.
\end{align}

For two bosonic fields $\phi(z)$ and $\chi(w)$,
Wick's theorem \cite{Wick1950} gives
\begin{align}
:e^{a\phi(z)}::e^{b\chi(w)}:
&=
e^{ab\contraction{}{\phi(z)}{}{\chi(w)}
\phi(z)\chi(w)}
:e^{a\phi(z)+b\chi(w)}:.
\label{dfkdtestse}
\end{align}

\subsection{Point splitting}

We shall use the point-splitting method of
\cite[Appendix~A, Eq.~(A.2)]{Camino1998}
to express the normal-ordered product of the derivative
of a bosonic field and a vertex operator as
\begin{align}
:\alpha'(z)\exp(\eta(z)):
&=
\lim_{z_1\to z}
\partial_{z_1}
:\exp\left((\eta-\alpha)(z)+\alpha(z_1)\right):.
\label{hohji}
\end{align}
Throughout this paper, normal ordering will be implicit,
and the coincident-point limit will be understood to be
taken after all indicated derivatives have been evaluated.
All operator-product expansions are understood as formal
local expansions in powers of $z-w$, with the fields
appearing in their coefficients evaluated at $w$.
\subsection{Fa\`a di Bruno polynomials}
We use the shorthand
\[
\partial\equiv\frac{\partial}{\partial w}.
\]

For $l\in\mathbb Z_{\geq0}$, define the Fa\`a di Bruno polynomials \cite{FaadiBruno1855} by
\begin{align}
P_l\equiv P_l(f)=e^{-f}\partial^l e^f. \label{faadibruno}
\end{align}
The definition gives
\begin{align}
P_0=1;~P_1=\partial f;~P_l=(P_1+\partial)^l 1;~P_{l+1}=(P_1+\partial)P_l. \label{recursivedk}
\end{align}

\subsubsection{Combinatorial identities}\label{identdkdfkdh} 
Throughout this paper, sums with unspecified bounds are taken
over all integers. A sum whose upper limit is smaller than its
lower limit is understood to be zero.

For $a\in \mathbb{C}$ and $k\in\mathbb Z$, we define
\[
\binom{a}{k}=
\begin{cases}
\displaystyle\frac{a(a-1)\cdots(a-k+1)}{k!}, & k>0,\\[6pt]
1, & k=0,\\
0, & k<0.
\end{cases}
\]

We also define the falling factorial
\[
(a)_{(k)}=
\begin{cases}
a(a-1)\cdots(a-k+1), & k>0,\\
1, & k=0,
\end{cases}
\]
for $a\in\mathbb{C}$ and $k\in\mathbb{Z}_{\geq0}$.

For nonnegative integers $l,n$, we have
\begin{align}
\sum_{s=0}^{l}(-1)^s{l \choose s}P_n \partial^s P_{l-s}
&=\sum_{s=0}^{l}(-1)^s{l \choose s}\partial^s P_{l+n-s}.
\label{kdjjdd}
\end{align}
This follows from the differential-operator identity of
Aratyn et al.~\cite[Eq.~(5.16)]{Aratyn1993}, by applying
both sides to the constant function $1$ and identifying
their $J$ with $\partial f$.

For $a\in\mathbb C$, $b\in\mathbb Z$ and
$n\in\mathbb Z_{\geq0}$, the Chu--Vandermonde identity is
\begin{align}
{a \choose b}
&=\sum_{j=0}^{n}{n \choose j}{a-n \choose b-j}.
\label{edujedhe}
\end{align}

\subsection{The $\mathcal J$ operators and their general OPE}
We now generalise the point-splitting method in
(\ref{hohji}). Let $\alpha,\beta\in\mathbb{C}^d$
and let $l$ be a non-negative integer.
We define the operator
\begin{align}
\mathcal{J}[l,\alpha,\beta](z)
&=P_l(-\alpha(z))\exp(\beta(z))\nonumber\\
&=\lim_{z_1\to z}
\partial_{z_1}^l
\exp\left((\alpha+\beta)(z)-\alpha(z_1)\right).
\end{align}
As throughout this paper, the exponential on the
right-hand side is understood to be normal ordered.

We now calculate the OPE of one $\mathcal{J}$ operator with another.  We write,
\begin{align*}
&\mathcal{J}[l,\alpha,\beta](z)\mathcal{J}[m,\gamma,\delta](w)\\
&=\lim_{z_1\rightarrow z} \partial^{l}_{z_1}\exp((\alpha+\beta)(z)-\alpha(z_1)) \lim_{w_1\rightarrow w} \partial^{m}_{w_1}\exp((\gamma+\delta)(w)-\gamma(w_1))\\
&=\lim_{z_1\rightarrow z, w_1\rightarrow w}\partial_{z_1}^{l}\partial_{w_1}^m\left(\exp((\alpha+\beta)(z)-\alpha(z_1))\exp((\gamma+\delta)(w)-\gamma(w_1))\right).
\end{align*}

Using Wick's theorem in the form (\ref{dfkdtestse}), we obtain	
\begin{align}
 \mathcal{J}[l,\alpha,\beta](z)\mathcal{J}[m,\gamma,\delta](w)&=\lim_{z_1\rightarrow z, w_1\rightarrow w}\partial_{z_1}^{l}\partial_{w_1}^m\left(
\frac{(z-w_1)^{(\alpha+\beta)\cdot\gamma} (z_1-w)^{\alpha\cdot(\gamma+\delta)}}{(z-w)^{(\alpha+\beta)\cdot(\gamma+\delta)}(z_1-w_1)^{\alpha\cdot\gamma}}\right. \nonumber\\
& \left. \exp((\alpha+\beta)(z)+(\gamma+\delta)(w)-\alpha(z_1)-\gamma(w_1))
\right).\label{dkzseydy}
\end{align}

Since the exponents involve dot products that need not be integers,
their derivatives are expressed using the generalised binomial
coefficients defined above.

\begin{lem}
The derivatives and limits in the preceding point-splitting expression can be evaluated to give the following finite sum for the product of the two $\mathcal{J}$ operators:
\begin{align}
\mathcal{J}[l,\alpha,\beta](z)\mathcal{J}[m,\gamma,\delta](w)=&\sum_{q=0}^l \sum_{r=0}^{m}\frac{(-1)^r l! m!}{(l-q)!(m-r)!}Y(\alpha\cdot\gamma,\alpha\cdot\delta,\beta\cdot\gamma,r,q)\nonumber\\
&P_{l-q}(-\alpha(z))P_{m-r}(-\gamma(w))\frac{\exp(\beta(z)+\delta(w))}{(z-w)^{\beta\cdot\delta+r+q}}\label{fkfridrdrwf},
\end{align}
where $Y$ can be expressed in either of two equivalent forms:
\begin{align}
Y(a,b,c,r,q)&=\sum_j {a+c \choose r-j}{-a \choose j}{b-j \choose q}\label{dkdkkisdk}\\
&=\sum_j {a+b \choose q-j}{-a \choose j}{c-j \choose r}.\label{fkfideus}
\end{align}
\end{lem}
\begin{pf}
We prove (\ref{dkdkkisdk}) by differentiating first with respect to \(w_1\) and then with respect to \(z_1\), following the approach of \cite{Camino1998}. We use Leibniz's rule in the form
\begin{align*}
\partial^l (f g)=\sum_{r=0}^l{l \choose r}\left[\partial^r f\right] \left[\partial^{l-r} g\right].
\end{align*}
We now rewrite (\ref{dkzseydy}) as
\begin{align*}
&\mathcal{J}[l,\alpha,\beta](z)\mathcal{J}[m,\gamma,\delta](w)\\
=&\lim_{z_1\rightarrow z}\frac{1}{(z-w)^{(\alpha+\beta)\cdot(\gamma+\delta)}}\partial_{z_1}^{l}\Bigl((z_1-w)^{\alpha\cdot(\gamma+\delta)}\\
& \lim_{w_1\rightarrow w}\partial_{w_1}^m\Bigl(
\frac{(z-w_1)^{(\alpha+\beta)\cdot\gamma} }{(z_1-w_1)^{\alpha\cdot\gamma}}
 \exp((\alpha+\beta)(z)+(\gamma+\delta)(w)-\alpha(z_1)-\gamma(w_1))
\Bigr)\Bigr)\\
=&\lim_{z_1\rightarrow z}\frac{1}{(z-w)^{(\alpha+\beta)\cdot(\gamma+\delta)}}\partial_{z_1}^{l}\Bigl((z_1-w)^{\alpha\cdot(\gamma+\delta)} \lim_{w_1\rightarrow w}\\
&\sum_{r=0}^m {m \choose r} \left[\partial^{r}_{w_1}\frac{(z-w_1)^{(\alpha+\beta)\cdot\gamma} }{(z_1-w_1)^{\alpha\cdot\gamma}}\right]\\
&\left[ \partial^{m-r}_{w_1}\exp((\alpha+\beta)(z)+(\gamma+\delta)(w)-\alpha(z_1)-\gamma(w_1))\right]\Bigr)\\
=&\sum_{r=0}^m {m \choose r}\lim_{z_1\rightarrow z}\frac{1}{(z-w)^{(\alpha+\beta)\cdot(\gamma+\delta)}}\partial_{z_1}^{l}\Bigl((z_1-w)^{\alpha\cdot(\gamma+\delta)} \lim_{w_1\rightarrow w}\\
& \sum_{j=0}^r {r \choose j}\left[\partial_{w_1}^{r-j}(z-w_1)^{(\alpha+\beta)\cdot\gamma} \right]\left[\partial_{w_1}^j(z_1-w_1)^{-\alpha\cdot\gamma}\right]\\
&P_{m-r}(-\gamma(w_1))\exp((\alpha+\beta)(z)+(\gamma+\delta)(w)-\alpha(z_1)-\gamma(w_1))\Bigr)\\
=&\sum_{r=0}^m \sum_{j=0}^r (-1)^r{m \choose r}{r \choose j}\lim_{z_1\rightarrow z}\frac{1}{(z-w)^{(\alpha+\beta)\cdot(\gamma+\delta)}}\partial_{z_1}^{l}\Bigl((z_1-w)^{\alpha\cdot(\gamma+\delta)} \\
& \lim_{w_1\rightarrow w}\left[((\alpha+\beta)\cdot\gamma)_{(r-j)}(z-w_1)^{(\alpha+\beta)\cdot\gamma-(r-j)} \right]\left[(-\alpha\cdot\gamma)_{(j)}(z_1-w_1)^{-\alpha\cdot\gamma-j}\right]\\
&P_{m-r}(-\gamma(w_1))\exp((\alpha+\beta)(z)+(\gamma+\delta)(w)-\alpha(z_1)-\gamma(w_1))\Bigr).
\end{align*}
We have applied Leibniz's rule twice to the $w_1$-derivatives.
The index $r$ counts the derivatives acting on the scalar power
factors, while $j$ counts those acting specifically on
$(z_1-w_1)^{-\alpha\cdot\gamma}$.
The remaining $m-r$ derivatives act on the exponential.

Taking the limit $w_1\rightarrow w$, we obtain
\begin{align*}
&\mathcal{J}[l,\alpha,\beta](z)\mathcal{J}[m,\gamma,\delta](w)\\
=&\sum_{r=0}^m \sum_{j=0}^r (-1)^r{m \choose r}r! {(\alpha+\beta)\cdot\gamma \choose r-j} {-\alpha\cdot\gamma \choose j} \frac{1}{(z-w)^{(\alpha+\beta)\cdot\delta+r-j}}\\
& P_{m-r}(-\gamma(w))
\lim_{z_1\rightarrow z}\partial_{z_1}^{l}
\left[
(z_1-w)^{\alpha\cdot\delta-j}
\exp((\alpha+\beta)(z)+\delta(w)-\alpha(z_1))
\right].
\end{align*}
We now apply Leibniz's rule to the $z_1$-derivatives.
Of the $l$ derivatives, $q$ act on the power of $z_1-w$,
while the remaining $l-q$ act on the exponential.
\begin{align*}
&\mathcal{J}[l,\alpha,\beta](z)\mathcal{J}[m,\gamma,\delta](w)\\
=&\sum_{r=0}^m \sum_{j=0}^r (-1)^r{m \choose r}r! {(\alpha+\beta)\cdot\gamma \choose r-j} {-\alpha\cdot\gamma \choose j} \frac{1}{(z-w)^{(\alpha+\beta)\cdot\delta+r-j}}\\
&\lim_{z_1\rightarrow z}\sum_{q=0}^{l}{l \choose q} [\partial_{z_1}^{q}(z_1-w)^{\alpha\cdot\delta-j}]P_{m-r}(-\gamma(w))[\partial_{z_1}^{l-q}\exp((\alpha+\beta)(z)+\delta(w)-\alpha(z_1))]\\
=&\sum_{r=0}^m \sum_{j=0}^r (-1)^r {m \choose r}r! {(\alpha+\beta)\cdot\gamma \choose r-j} {-\alpha\cdot\gamma \choose j} \frac{1}{(z-w)^{(\alpha+\beta)\cdot\delta+r-j}}\\
&\lim_{z_1\rightarrow z}\sum_{q=0}^{l}{l \choose q} [(\alpha\cdot\delta-j)_{(q)}(z_1-w)^{\alpha\cdot\delta-j-q}]P_{l-q}(-\alpha(z_1)) P_{m-r}(-\gamma(w))\\
&\exp((\alpha+\beta)(z)+\delta(w)-\alpha(z_1))\\
=&\sum_{q=0}^{l}\sum_{r=0}^m \sum_{j=0}^r (-1)^r {m \choose r}{l \choose q}q!r! {(\alpha+\beta)\cdot\gamma \choose r-j} {-\alpha\cdot\gamma \choose j}{\alpha\cdot\delta-j \choose q} \frac{1}{(z-w)^{\beta\cdot\delta+r+q}}\\
&P_{l-q}(-\alpha(z)) P_{m-r}(-\gamma(w))\exp(\beta(z)+\delta(w)),
\end{align*}
This proves (\ref{dkdkkisdk}). The alternative form (\ref{fkfideus}) follows by reversing the order of differentiation.
\end{pf}

To express (\ref{fkfridrdrwf}) as an OPE in powers of $z-w$,
we Taylor expand the remaining fields at $z$ about $w$:
\begin{align*}
P_{l-q}(-\alpha(z)) &=\sum_{n=0}^{\infty} \frac{(z-w)^n}{n!} \partial^{n} P_{l-q}(-\alpha(w)).
\end{align*}
Similarly, Taylor expanding the exponential about $w$ yields an infinite series of Fa\`a di Bruno polynomials:
\begin{align*}
\exp(\beta(z)+\delta(w))&=\sum_{n^{\prime}=0}^{\infty}P_{n^{\prime}}(\beta(w))\exp((\beta+\delta)(w))\frac{(z-w)^{n^{\prime}}}{n^{\prime}!}.
\end{align*}
Substituting these Taylor expansions gives the OPE,
with all coefficient fields evaluated at $w$:
\begin{align}
&\mathcal{J}[l,\alpha,\beta](z)\mathcal{J}[m,\gamma,\delta](w)\nonumber\\
&=\sum_{n,n^{\prime}=0}^{\infty}\sum_{q=0}^l \sum_{r=0}^{m}\frac{(-1)^r l! m!}{(l-q)!(m-r)!n!n^{\prime}!}Y(\alpha\cdot\gamma,\alpha\cdot\delta,\beta\cdot\gamma,r,q)\nonumber\\
&\partial^n P_{l-q}(-\alpha(w))P_{m-r}(-\gamma(w))P_{n^{\prime}}(\beta(w))\frac{\exp((\beta+\delta)(w))}{(z-w)^{\beta\cdot\delta+r+q-n-n^{\prime}}}.\label{gfhfh}
\end{align}

\section{Convolution of Faà di Bruno polynomials}

\subsection{The convolution identity}
In the previous section, we derived the general OPE of two
$\mathcal{J}$ operators. We now show that, under certain
conditions, the leading term takes a particularly simple form:
the two Fa\`a di Bruno indices combine additively.
\label{sksjjkmawjq}
\begin{thm}\label{mainresult}
Let $l,m\in\mathbb Z_{\geq0}$, and suppose that
\begin{align}
\alpha=\gamma;~~~\alpha\cdot\alpha=-1;~~~
\alpha\cdot\beta=-l;~~~\alpha\cdot\delta=-m.
\label{requiredjkdfj}
\end{align}
Then the OPE (\ref{gfhfh}) satisfies
\begin{align}
\mathcal{J}[l,\alpha,\beta](z)\mathcal{J}[m,\gamma,\delta](w)
&=\frac{\mathcal{J}[l+m,\alpha,\beta+\delta](w)}
{(z-w)^{\beta\cdot\delta}}+\dots
\label{dudlasyaisk}
\end{align}
The omitted terms are of higher order in $z-w$.
\end{thm}
The additive property expressed by Theorem \ref{mainresult}
is relevant to the parafermionic constructions of
Bilal \cite{Bilal1989} and Camino et al.
\cite{Camino1998}. The latter authors state the
corresponding polynomial identity explicitly in
Appendix A, equation (A.14). We establish this
identity below and then use it to complete the
proof of Theorem \ref{mainresult}.

We prove Theorem \ref{mainresult} by considering separately the cases in which at least one of $l,m$ is zero and the case $l,m\geq 1$. The boundary cases follow directly from the general OPE derived in the previous section. For $l,m\geq 1$, the proof requires a combinatorial identity involving Faà di Bruno polynomials, which we establish below.

\begin{lem}\label{boundarycases}
Theorem \ref{mainresult} holds when $l=0$ or $m=0$.
\end{lem}

\begin{pf}
We consider separately the cases \(m=0\) and \(l=0\).

When $m=0$, the sum over $r$ reduces to the term $r=0$.
We have
\begin{align*}
Y(-1,0,-l,0,q)
&=\sum_j
\binom{-1-l}{-j}
\binom{1}{j}
\binom{-j}{q}\\
&=\binom{-1-l}{0}\binom{1}{0}\binom{0}{q}\\
&=\delta_{q,0}.
\end{align*}
In the second equality, we have used the fact that
$\binom{-1-l}{-j}=0$ for $j>0$, while
$\binom{1}{j}=0$ for $j<0$. Thus only $j=0$
contributes.

Consequently, only $q=0$ contributes to the general
OPE in Lemma 1. Expanding the remaining fields about
$w$, we obtain
\begin{align*}
&\mathcal J[l,\alpha,\beta](z)
\mathcal J[0,\alpha,\delta](w)\\
&\qquad=
\frac{\mathcal J[l,\alpha,\beta+\delta](w)}
{(z-w)^{\beta\cdot\delta}}
+\cdots.
\end{align*}

Similarly, when $l=0$, the sum over $q$ reduces to
the term $q=0$. We have
\begin{align*}
Y(-1,-m,0,r,0)
&=\sum_j
\binom{-1}{r-j}
\binom{1}{j}
\binom{-m-j}{0}\\
&=\binom{-1}{r}+\binom{-1}{r-1}\\
&=\delta_{r,0}.
\end{align*}
Here only $j=0,1$ can contribute. For $r\geq1$,
the two terms cancel because
\[
\binom{-1}{r}=(-1)^r,
\qquad
\binom{-1}{r-1}=(-1)^{r-1}.
\]
For $r=0$, the coefficient is $1$.

Consequently, only $r=0$ contributes to the general
OPE in Lemma 1. Expanding the remaining fields about
$w$, we obtain
\begin{align*}
&\mathcal J[0,\alpha,\beta](z)
\mathcal J[m,\alpha,\delta](w)\\
&\qquad=
\frac{\mathcal J[m,\alpha,\beta+\delta](w)}
{(z-w)^{\beta\cdot\delta}}
+\cdots.
\end{align*}

These arguments also cover the case $l=m=0$.
\end{pf}

It remains to establish the result when $l,m\geq1$.

In this case, using (\ref{fkfideus}), we obtain
\begin{align*}
Y(-1,-m,-l,r,q)
&=\sum_j\binom{-m-1}{q-j}\binom{1}{j}
\binom{-l-j}{r}\\
&=\binom{-m-1}{q}\binom{-l}{r}
+\binom{-m-1}{q-1}\binom{-l-1}{r}\\
&=(-1)^{r+q}
\frac{(l+r-1)!(m+q-1)!}{r!q!l!m!}(ml-qr),
\end{align*}
and the sum in (\ref{gfhfh}) becomes
\begin{align*}
&\mathcal{J}[l,\alpha,\beta](z)\mathcal{J}[m,\alpha,\delta](w)\\
=&\sum_{n^{\prime}=0}^{\infty}\sum_{q=0}^{l}\sum_{r=0}^{m}\sum_{n=0}^{\infty}(-1)^q(ml-qr)\frac{(l+r-1)!(m+q-1)!}{r!q!(m-r)!(l-q)!n!n^{\prime}!}\\
&\partial^{n}P_{l-q}(-\alpha (w))P_{m-r}(-\alpha (w))P_{n^{\prime}}(\beta(w))\frac{\exp((\beta+\delta)(w))}{(z-w)^{\beta\cdot\delta+r+q-n-n^{\prime}}}\\
=&\sum_{n^{\prime}=0}^{\infty}\sum_{q=0}^{l}\sum_{r=0}^{m}\sum_{n=-r-q}^{\infty}(-1)^q\frac{(m+q-1)!(l+r-1)!}{r!q!(m-r)!(l-q)!(r+q+n)!n^{\prime}!}(ml-qr)\\
&\partial^{r+q+n}P_{l-q}(-\alpha (w))P_{m-r}(-\alpha (w))P_{n^{\prime}}(\beta(w))\frac{\exp((\beta+\delta) (w))}{(z-w)^{\beta\cdot\delta-n-n^{\prime}}},
\end{align*}
In the last equality, we have replaced the original summation index $n$ by $r+q+n.$ Since the original index was nonnegative, the new index runs from $-r-q$ to infinity.

We collect powers of $(z-w)$ using the Heaviside function
$H(k)=1$ for $k\geq 0$ and $H(k)=0$ otherwise.

Whenever a summand contains Heaviside factors restricting its indices, the summand is understood to vanish outside the range on which all those factors equal \(1\). All factorial expressions are therefore evaluated only on that range.

\begin{align}
&\mathcal{J}[l,\alpha,\beta](z)\mathcal{J}[m,\alpha,\delta](w)\nonumber\\
=&\sum_{n^{\prime}=0}^{\infty}\sum_{n \in \mathbb{Z}}\left(\sum_{r=0}^{m}\sum_{q=0}^{l}\frac{H(r+q+n)}{(r+q+n)!}(-1)^q\frac{(m+q-1)!(l+r-1)!}{r!q!(m-r)!(l-q)!}(ml-qr)\right.\nonumber\\
&\left.\partial^{r+q+n}P_{l-q}(-\alpha (w))P_{m-r}(-\alpha (w))\frac{\exp((\beta+\delta)(w))}{(z-w)^{\beta\cdot\delta}}\right)\nonumber\\
&\frac{P_{n^{\prime}}(\beta(w))(z-w)^{n+n^{\prime}}}{n^{\prime}!}.\label{dsedgasgtsdkv}
\end{align} 

We now collect the terms in (\ref{dsedgasgtsdkv})
according to their powers of $z-w$.
For each integer $n$, let $G(l,m,n)$ denote the
coefficient of $(z-w)^{n-\beta\cdot\delta}$
in the part of the expansion with $n'=0$,
after factoring out $\exp((\beta+\delta)(w))$:

\begin{align}
G(l,m,n)=&\sum_{r=0}^{m}\sum_{q=0}^{l} \frac{H(r+q+n)}{(r+q+n)!}\nonumber\\
&(-1)^q\frac{(m+q-1)!(l+r-1)!(ml-qr)}{r!q!(m-r)!(l-q)!}P_{m-r}\partial^{r+q+n}P_{l-q}\nonumber\\
=&\sum_{r=0}^{m}\sum_{q=0}^{l} \frac{H(r+q+n)}{(r+q+n)!}\nonumber\\
&(-1)^q\frac{(m+q-1)!(l+r-1)!(m(l+r)-(m+q)r)}{r!q!(m-r)!(l-q)!}P_{m-r}\partial^{r+q+n}P_{l-q}\nonumber\\
=&\sum_{r=0}^{m}\sum_{q=0}^{l} \frac{H(r+q+n)}{(r+q+n)!}\nonumber\\
&(-1)^q\left[\frac{m(m+q-1)!(l+r)!}{r!q!(m-r)!(l-q)!}-\frac{r(m+q)!(l+r-1)!}{r!q!(m-r)!(l-q)!}\right]\nonumber\\
&P_{m-r}\partial^{r+q+n}P_{l-q},\label{seksyhawj}
\end{align}
Here and throughout the remainder of the proof, we suppress
the common argument $-\alpha(w)$ of the Fa\`a di Bruno
polynomials. Thus $P_k$ denotes $P_k(-\alpha(w))$,
and derivatives act with respect to $w$.

We first introduce $s=r+q$, and then replace $r$
with $m-r$. Under these substitutions, the polynomial
factor $P_{m-r}\partial^{r+q+n}P_{l-q}$ becomes
$P_r\partial^{s+n}P_{l+m-r-s}$.
The original bounds $0\leq r\leq m$ and $0\leq q\leq l$
become $0\leq r\leq m$ and
$m-r\leq s\leq l+m-r$.
Expressing the factorial coefficients as binomial
coefficients gives
\begin{align}
G(l,m,n)=&\sum_{r=0}^{m}\sum_{s=m-r}^{\infty} (-1)^{s+m+r} H(s+n)H(r)H(l+m-r-s)\nonumber\\
&\left[\frac{(l+m-r)!}{(s+n)!(l+m-r-s)!}{s+r-1 \choose m-1}{m \choose r}\right.\nonumber\\
&\left.-\frac{l!m!}{(l+m+n-r)!r!}{l+m-r-1 \choose m-r-1}{l+m+n-r \choose s+n}{s+r \choose m}\right]\nonumber\\
&P_{r}\partial^{s+n}P_{l+m-r-s}.\label{stgdgtdg}
\end{align}
Here the summation bounds have been extended using the convention that \(\binom{a}{k}=0\) whenever \(k<0\). In particular, \(\binom mr\) restricts \(r\) to \(0\leq r\leq m\), while \(H(l+m-r-s)\) enforces \(s\leq l+m-r\).

In fact we can widen the bounds even more than this.
Observe
\begin{align*}
s<-n~&\implies H(s+n)=0\\
0<s<m-r~&\implies \left({s+r-1\choose m-1}=0\text{ or }{m \choose r}=0\right)\\
&\text{ and }\left(H(r)=0 \text{ or }{s+r \choose m}=0\right).
\end{align*}
Since we are interested only in the leading singular term of equation (\ref{dsedgasgtsdkv}), it is sufficient to consider \(n\leq 0\). For these values of \(n\), we may extend the summation over \(s\) to all integers outside the set \(\mathfrak X=\{s\in\mathbb Z:-n\leq s\leq0\}\). The only contribution removed from the original summation is \(P_lP_m\) when \(n=0\). Thus,
\begin{align}
G(l,m,n)=&\sum_{r}\sum_{s\in \mathbb{Z} \backslash \mathfrak{X}} (-1)^{s+m+r} H(s+n)H(r)H(l+m-r-s)\nonumber\\
&\left[\frac{(l+m-r)!}{(s+n)!(l+m-r-s)!}{s+r-1 \choose m-1}{m \choose r}\right.\nonumber\\
&\left.-\frac{l!m!}{(l+m+n-r)!r!}{l+m-r-1 \choose m-r-1}{l+m+n-r \choose s+n}{s+r \choose m}\right]\nonumber\\
&P_{r}\partial^{s+n}P_{l+m-r-s}\nonumber\\
&+\delta_{n,0}P_lP_m,\label{sdifdg}
\end{align}

For the remainder of the convolution argument, we assume
that $l,m$ are positive integers and $n$ is a nonpositive
integer.

To facilitate the subsequent combinatorial manipulations,
we introduce the auxiliary expression $G(l,m,n)^*$,
obtained by extending the summation over $s$ to all integers.
We will show that $G(l,m,n)^*=0$, and then recover
$G(l,m,n)$ by accounting for the excluded values
$s\in\mathfrak X$.
\begin{align}
G(l,m,n)^{*}=&\sum_{r}\sum_{s} (-1)^{s+m+r} H(s+n)H(r)H(l+m-r-s)\nonumber\\
&\left[\frac{(l+m-r)!}{(s+n)!(l+m-r-s)!}{s+r-1 \choose m-1}{m \choose r}\right.\nonumber\\
&\left.-\frac{l!m!}{(l+m+n-r)!r!}{l+m-r-1 \choose m-r-1}{l+m+n-r \choose s+n}{s+r \choose m}\right]\nonumber\\
&P_{r}\partial^{s+n}P_{l+m-r-s}.\label{girtdedet}
\end{align}
Change variables so that $s+n=t$:
\begin{align}
G(l,m,n)^{*}=&\sum_{r}\sum_{t}(-1)^{t+n+m+r} H(t)H(r)H(l+m+n-r-t)\nonumber\\
&\left[\frac{(l+m-r)!}{t!(l+m+n-r-t)!}{t+r-n-1 \choose m-1}{m \choose r}\right.\nonumber\\
&\left.-\frac{l!m!}{r!(l+m+n-r)!}{l+m-r-1 \choose m-r-1}{l+m+n-r \choose t}{t+r-n \choose m}\right]\nonumber\\
&P_{r}\partial^{t}P_{l+m+n-r-t}.\label{stgdaqg}
\end{align}

\begin{lem}
Equation (\ref{stgdaqg}) can be written as
\begin{align}
&G(l,m,n)^*\nonumber\\
=&{}
\sum_{r=0}^{m}
\sum_{j=0}^{\min(m-1,l+m+n-r)}
(-1)^{r+m+n+j}\nonumber\\
&\quad\times
\frac{(l+m-r)!}{(l+m+n-r)!}
\binom{m}{r}
\binom{l+m-j-1}{m-j-1}
\binom{l+m+n-r}{j}
\nonumber\\
&\qquad\times
\sum_{t=0}^{l+m+n-r-j}
(-1)^t
\binom{l+m+n-r-j}{t}
P_r\partial^t P_{l+m+n-r-t}
\nonumber\\[6pt]
&-
\sum_{j=0}^{m}
\sum_{r=0}^{\min(m-1,l+m+n-j)}
(-1)^{r+m+n+j}\nonumber\\
&\quad\times
\frac{(l+m-j)!}{(l+m+n-j)!}
\binom{m}{j}
\binom{l+m-r-1}{m-r-1}
\binom{l+m+n-j}{r}
\nonumber\\
&\qquad\times
\sum_{t=0}^{l+m+n-r-j}
(-1)^t
\binom{l+m+n-r-j}{t}
P_r\partial^t P_{l+m+n-r-t}.
\label{dfusetsw}
\end{align}
\end{lem}

\begin{pf}
To prove the lemma, we rewrite the binomial coefficients
in (\ref{stgdaqg}) using the Chu--Vandermonde identity.
This will allow us to express $G(l,m,n)^*$ as a combination
of sums to which identity (\ref{kdjjdd}) can be applied.
We introduce two auxiliary quantities, $A$ and $B$,
containing the coefficients to be rearranged together
with their Heaviside factors. The remaining factors
will be displayed explicitly.

In the calculations below, the summation limits are
chosen so that all factorials are defined.

For convenience, define
\begin{align*}
A&=
H(r)H(t)H(l+m+n-r-t)
\frac{1}{(l+m+n-r-t)!t!}
\binom{t+r-n-1}{m-1},\\[4pt]
B&=
H(r)H(t)H(l+m+n-r-t)
\frac{1}{(l+m+n-r)!}
\binom{l+m+n-r}{t}
\binom{t+r-n}{m}.
\end{align*}
Here the Heaviside factors specify the support
\[
r\geq 0,\qquad t\geq 0,\qquad r+t\leq l+m+n.
\]
Both $A$ and $B$ are defined to be zero outside this range.

Then
\begin{align}
G(l,m,n)^*
=&\sum_{r=0}^{m}
\sum_{t=0}^{l+m+n-r}
(-1)^{m+r+n+t}
\Bigg[
(l+m-r)!\binom{m}{r}A \nonumber\\
&\qquad\qquad
-\frac{l!m!}{r!}
\binom{l+m-r-1}{m-r-1}B
\Bigg]
P_r\partial^tP_{l+m+n-r-t}.
\label{gstarreworked}
\end{align}

We calculate $A$ on the support of its Heaviside factors.
On this support, all three Heaviside factors equal one, and
\begin{align*}
A
=&\frac{1}{(l+m+n-r-t)!t!}
\binom{t+r-n-1}{m-1}\\[4pt]
=&(-1)^{m-1}
\frac{1}{(l+m+n-r-t)!t!}
\binom{m+n-r-t-1}{m-1}\\[4pt]
=&(-1)^{m-1}
\frac{1}{(l+m+n-r-t)!t!}
\sum_{j=0}^{\min(m-1,l+m+n-r-t)}
\binom{-l-1}{m-j-1}
\binom{l+m+n-r-t}{j}\\[4pt]
=&\frac{1}{(l+m+n-r-t)!t!}
\sum_{j=0}^{\min(m-1,l+m+n-r-t)}
(-1)^j
\binom{l+m-j-1}{m-j-1}
\binom{l+m+n-r-t}{j}\\[4pt]
=&
\sum_{j=0}^{\min(m-1,l+m+n-r-t)}
(-1)^j
\frac{1}{(l+m+n-r-j-t)!j!t!}
\binom{l+m-j-1}{m-j-1}\\[4pt]
=&
\sum_{j=0}^{\min(m-1,l+m+n-r-t)}
\frac{(-1)^j}{(l+m+n-r)!}
\binom{l+m-j-1}{m-j-1}\\
&\qquad\times
\binom{l+m+n-r}{j}
\binom{l+m+n-r-j}{t}.
\end{align*}

We similarly calculate $B$ on the same support:
\begin{align*}
B
=&\frac{1}{(l+m+n-r)!}
\binom{l+m+n-r}{t}
\binom{t+r-n}{m}\\[4pt]
=&(-1)^m
\frac{1}{(l+m+n-r)!}
\binom{l+m+n-r}{t}
\binom{m-t+n-r-1}{m}\\[4pt]
=&(-1)^m
\frac{1}{(l+m+n-r)!}
\binom{l+m+n-r}{t}
\sum_{j=0}^{\min(m,l+m+n-r-t)}
\binom{-l-1}{m-j}
\binom{l+m+n-r-t}{j}\\[4pt]
=&(-1)^m\frac{1}{t!}
\sum_{j=0}^{\min(m,l+m+n-r-t)}
\binom{-l-1}{m-j}
\frac{1}{j!(l+m+n-r-t-j)!}\\[4pt]
=&\frac{1}{t!}
\sum_{j=0}^{\min(m,l+m+n-r-t)}
(-1)^j
\binom{l+m-j}{m-j}
\frac{1}{j!(l+m+n-r-t-j)!}\\[4pt]
=&
\sum_{j=0}^{\min(m,l+m+n-r-t)}
\frac{(-1)^j}{j!(l+m+n-r-j)!}
\binom{l+m-j}{m-j}
\binom{l+m+n-r-j}{t}.
\end{align*}

Substituting our expressions for $A$ and $B$ into
equation \eqref{gstarreworked}, we obtain
\begin{align*}
G(l,m,n)^*
={}&
\sum_{r=0}^{m}
\sum_{t=0}^{l+m+n-r}
\sum_{j=0}^{\min(m-1,l+m+n-r-t)}
(-1)^{m+r+n+t+j}\\
&\quad\times
\frac{(l+m-r)!}{(l+m+n-r)!}
\binom{m}{r}
\binom{l+m-j-1}{m-j-1}
\binom{l+m+n-r}{j}\\
&\quad\times
\binom{l+m+n-r-j}{t}
P_r\partial^tP_{l+m+n-r-t}\\
&-
\sum_{r=0}^{m-1}
\sum_{t=0}^{l+m+n-r}
\sum_{j=0}^{\min(m,l+m+n-r-t)}
(-1)^{m+r+n+t+j}\\
&\quad\times
\frac{(l+m-j)!}{(l+m+n-j)!}
\binom{m}{j}
\binom{l+m-r-1}{m-r-1}
\binom{l+m+n-j}{r}\\
&\quad\times
\binom{l+m+n-r-j}{t}
P_r\partial^tP_{l+m+n-r-t}.
\end{align*}

In the first term, we interchange the order of the
$t$- and $j$-summations. This is justified because
\[
\begin{aligned}
&0\leq t\leq l+m+n-r,\qquad
0\leq j\leq\min(m-1,l+m+n-r-t)\\
&\quad\Longleftrightarrow\quad
0\leq j\leq\min(m-1,l+m+n-r),\qquad
0\leq t\leq l+m+n-r-j.
\end{aligned}
\]

In the second term, we similarly interchange the
$t$- and $j$-summations, using
\[
\begin{aligned}
&0\leq t\leq l+m+n-r,\qquad
0\leq j\leq\min(m,l+m+n-r-t)\\
&\quad\Longleftrightarrow\quad
0\leq j\leq\min(m,l+m+n-r),\qquad
0\leq t\leq l+m+n-r-j.
\end{aligned}
\]
Since the second term has $0\leq r\leq m-1$,
we may then interchange the $r$- and $j$-summations.
The resulting summation limits are
\[
0\leq j\leq m,\qquad
0\leq r\leq\min(m-1,l+m+n-j).
\]

The resulting expression is precisely
(\ref{dfusetsw}), completing the proof.
\end{pf}

\begin{lem}
\begin{align}
G(l,m,n)^{*}=0.
\end{align}
\end{lem}

\begin{pf}
Equation (\ref{dfusetsw}) is useful, since it shows the first and second term as being nearly identical to each other under interchange of $j$ and $r$. 

We use (\ref{kdjjdd}) to show that these two terms are in fact equal and therefore cancel.
\begin{align*}
G(l,m,n)^{*}
=&\sum_{r=0}^{m}
\sum_{j=0}^{\min(m-1,l+m+n-r)}
(-1)^{r+m+n+j}\\
&\frac{(l+m-r)!}{(l+m+n-r)!}
{m \choose r}
{l+m-j-1 \choose m-j-1}
{l+m+n-r \choose j}\\
&\sum_{t=0}^{l+m+n-r-j}
(-1)^t
{l+m+n-r-j \choose t}
P_rP_j\partial^t P_{l+m+n-r-j-t}\\[6pt]
&-
\sum_{j=0}^{m}
\sum_{r=0}^{\min(m-1,l+m+n-j)}
(-1)^{r+m+n+j}\\
&\frac{(l+m-j)!}{(l+m+n-j)!}
{m \choose j}
{l+m-r-1 \choose m-r-1}
{l+m+n-j \choose r}\\
&\sum_{t=0}^{l+m+n-r-j}
(-1)^t
{l+m+n-r-j \choose t}
P_rP_j\partial^t P_{l+m+n-r-j-t}.
\end{align*}
In the second sum, interchange the indices \(r\) and \(j\). The resulting summand is identical to that of the first sum, since \(P_rP_j=P_jP_r\). The two sums therefore cancel, giving $G(l,m,n)^{*}=0$.
\end{pf}

\begin{thm}\label{convolutiontheorem}
Let $l,m$ be positive integers and let $n$ be
a nonpositive integer. Let $G(l,m,n)$ be defined
by (\ref{seksyhawj}). Then
\begin{align}
G(l,m,n)=
\begin{cases}
0, & n<0,\\
P_{l+m}, & n=0.
\end{cases}
\end{align}
\end{thm}

\begin{pf}
We now compare $G(l,m,n)$ with $G(l,m,n)^*$. By their definitions, the former differs from the latter by the exclusion of the values $s\in\mathfrak{X}$ from the double sum, together with the additional term $\delta_{n,0}P_lP_m$. Thus,
\begin{align}
G(l,m,n)&=G(l,m,n)^{*}+\delta_{n,0}P_lP_m-\sum_{r=0}^{m}\sum_{s\in \mathfrak{X}} (-1)^{s+m+r}\nonumber\\
&\left[\frac{(l+m-r)!}{(s+n)!(l+m-r-s)!}{s+r-1 \choose m-1}{m \choose r}\right.\nonumber\\
&\left.-\frac{l!m!}{(l+m+n-r)!r!}
{l+m-r-1\choose m-r-1}
{l+m+n-r\choose s+n}{s+r\choose m}
\right]
P_{r}\partial^{s+n}P_{l+m-r-s}\nonumber\\
&=\sum_{r=0}^{m}\sum_{s\in \mathfrak{X}} (-1)^{s+m+r} \nonumber\\
&\left[\frac{l!m!}{(l+m+n-r)!r!}{l+m-r-1 \choose m-r-1}{l+m+n-r \choose s+n}{s+r \choose m}\right.\nonumber\\
&\left.-\frac{(l+m-r)!}{(s+n)!(l+m-r-s)!}{s+r-1 \choose m-1}{m \choose r}\right]
P_{r}\partial^{s+n}P_{l+m-r-s}\nonumber\\
&+\delta_{n,0}P_lP_m.\label{highgfhft}
\end{align}
\par\medskip\noindent\textit{Case 1: $n<0$.}

Here $\mathfrak{X}=\emptyset$ and $\delta_{n,0}=0$.
Equation (\ref{highgfhft}) therefore gives
\[
G(l,m,n)=G(l,m,n)^*=0.
\]

\par\medskip\noindent\textit{Case 2: $n=0$.}

In this case,
\[
\mathfrak{X}=\{-n\le s\le 0\}=\{0\}.
\]

We break this into three terms:
\begin{align}
G(l,m,0)&=\mathfrak{A}(l,m)+\mathfrak{B}(l,m)+P_lP_m\nonumber\\
\mathfrak{A}(l,m)&=-\sum_{r=0}^{m}(-1)^{m+r} {r-1 \choose m-1}{m \choose r}P_{r}P_{l+m-r}\nonumber\\
\mathfrak{B}(l,m)&=\sum_{r=0}^{m}(-1)^{m+r}\frac{l!m!}{(l+m-r)!r!}{l+m-r-1 \choose m-r-1}{r \choose m}P_{r}P_{l+m-r}.\label{dkduseshgkqjufkf}
\end{align}

In $\mathfrak{A}$, the factor ${r-1\choose m-1}$
vanishes for $1\le r<m$. The terms $r=0$ and
$r=m$ contribute $P_{l+m}$ and $-P_lP_m$,
respectively.

In $\mathfrak{B}$, the factor ${r\choose m}$
vanishes for $r<m$, while at $r=m$ the factor
${l+m-r-1\choose m-r-1}$ vanishes.
Hence,
\begin{align}
\mathfrak{A}(l,m)&=P_{l+m}-P_lP_m\nonumber\\
\mathfrak{B}(l,m)&=0\nonumber\\
G(l,m,0)&=\mathfrak{A}(l,m)+\mathfrak{B}(l,m)+P_lP_m=P_{l+m}\label{sdkdfidkdgd}
\end{align}
\end{pf}

\begin{pf}
We can now complete the proof of Theorem \ref{mainresult}.
By Lemma \ref{boundarycases}, it remains only to consider
$l,m\geq 1$.

Returning to equation (\ref{dsedgasgtsdkv}), we write
\begin{align*}
&\mathcal{J}[l,\alpha,\beta](z)\mathcal{J}[m,\alpha,\delta](w)\\
=&\sum_{n'=0}^{\infty}\sum_{n\in\mathbb{Z}}
G(l,m,n)
\frac{P_{n'}(\beta(w))(z-w)^{n+n'}
\exp((\beta+\delta)(w))}
{n'!(z-w)^{\beta\cdot\delta}}\\
=&\frac{P_{l+m}(-\alpha(w))
\exp((\beta+\delta)(w))}
{(z-w)^{\beta\cdot\delta}}+\dots\\
=&\frac{\mathcal{J}[l+m,\alpha,\beta+\delta](w)}
{(z-w)^{\beta\cdot\delta}}+\dots.
\end{align*}
In the second equality, we have applied
Theorem \ref{convolutiontheorem}, which gives
$G(l,m,n)=0$ for $n<0$ and
$G(l,m,0)=P_{l+m}$.
The ellipsis denotes higher order terms, arising from $n+n'>0$.
The theorem is thus proved.
\end{pf}

\section{Conclusion}

We have presented a detailed proof of the
Fa\`a di Bruno convolution identity stated by
Camino, Ramallo and S\'anchez de Santos
\cite{Camino1998}. The argument establishes
both the cancellation of the potentially more
singular contributions and the additive combination
of polynomial indices in the leading term.

The proof retains the finite-sum and combinatorial
approach of the original parafermion calculation.
It does not address mixed-sign products, which
require a separate analysis.

\section*{Acknowledgements}

The author would like to thank his PhD supervisors,
Yao-Zhong Zhang and Wen-Li Yang, for their guidance
and support during his doctoral studies at
The University of Queensland.

\section*{Conflict of Interest}

The author has no conflicts to disclose.

\section*{Data Availability}

Data sharing is not applicable to this article as no new data were created or analyzed in this study.

\end{document}